Review

# Ionic thermoelectric materials and devices

Dan Zhao[a,*], Alois Würger[b], Xavier Crispin[a,c]

[a]*Laboratory of Organic Electronics, Linkoping University, Norrkoping 60247, Sweden*
[b]*Univ. Bordeaux & CNRS, LOMA (UMR 5798), F-33405 Talence, France*
[c]*Wallenberg Wood Science Center, ITN, Linköping University, Norrköping, Sweden*

[*]Corresponding author.
*E-mail address:* dan.zhao@liu.se (D. Zhao)

**ABSTRACT**
The tremendous amount of wasted heat from solar radiation and industry dissipation has motivated the development of thermoelectric concepts that directly convert heat into electricity. The main challenge in practical applications for thermoelectrics is the high cost from both materials and manufacturing. Recently, breakthrough progresses in ionic thermoelectrics open up new possibilities to charge energy storage devices when submitted to a temperature gradient. The charging voltage is internally from the ionic Seebeck effect of the electrolyte between two electrodes. Hence electrolytes with high thermoelectric figure of merit are classified as ionic thermoelectric materials. Most ionic thermoelectric materials are composed of abundant elements, and they can generate hundreds of times larger thermal voltage than that of electronic materials. This emerging thermoelectric category brings new hope to fabricate low cost and large area heat-to-energy conversion devices, and triggers a renewed interest for ionic thermodiffusion. In this review, we summarize the state of the art in the new field of ionic thermoelectrics, from the driving force of the ionic thermodiffusion to material and application developments. We present a general map of ionic thermoelectric materials, discuss the unique characters of each type of the reported electrolytes, and propose potential optimization and future topics of ionic thermoelectrics.

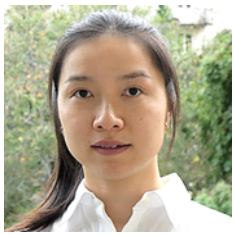

**Dan Zhao** is an Assistant professor at the Laboratory of Organic Electronics (LOE) at Linköping University. She received her B.S. from the Chemistry department in Nankai University in 2008. After finishing Ph.D. in Renmin University in China, she joined LOE as a post-doctoral researcher in 2013. Her research interest includes ionic thermoelectrics, electrolyte based and dielectric materials.

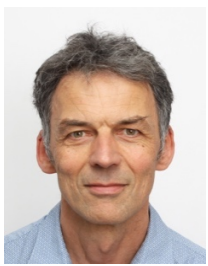

**Alois Würger** is a Professor in the Condensed Matter Theory group of LOMA, at Université de Bordeaux. He received his Ph.D from University of Erlangen, and his venia legend from University Heidelberg, and was a Research associate at the Institut Laue-Langevin in Grenoble, where he worked on low-temperature physics and quantum dissipation. His present research interests include non-equilibrium complex fluids, colloidal thermophoresis, nanofluidics and thermoelectric effects.

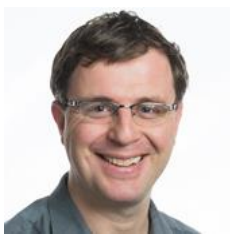

**Xavier Crispin** is a Professor at the Laboratory of Organic Electronics (LOE) in Linköping University. He obtained his B.S. and Ph.D. in University of Mons, Belgium, and developed his own research group at Linköping University as Marie-Curie and ERC fellow. He was awarded the Tage Erlander and Göran Gustafsson prizes by the Swedish Academy of Science for his work on organic thermoelectrics. His research interest is the use of organic energy materials in batteries, supercapacitors, thermoelectrics and fuel cells.

# 1. Introduction

The energy crisis is one of the most critical issues for modern society. The majority of the world's power is generated by heat engines with efficiency of only 30%–40% [1,2], which means that at least 60% of the consumed fossil fuel dissipates typically in the form of waste heat. Threatened by the gradually depleting fossil fuel, the exploration for renewable energy conversion technologies has attracted intensive attention. Sustainable solar radiation could cover thousands of times of the energy need for the whole world and has been considered as the most promising future energy source. Up to now, low-cost solar cells can only convert approximately 20% of the solar radiation into electricity [3], whereas 80% of the energy in the form of heat remains unused. The large amount of waste heat has attracted a strong interest in heat-to-electricity conversion to contribute in powering our society. In this context, thermoelectric effect that can directly convert a heat flow into electricity is the dream solution if it could be efficient enough.

Thermoelectric phenomena are related to the interplay between electrical and heat transport, thus going beyond the Ohm's law and the Fourier's law. Thermoelectric effects can be viewed as the result of mutual interaction between the two transport processes. The strength of the interaction in a material can be descibed by the Seebeck coefficient ($\alpha$), which is defined as the ratio between the generated open voltage and the temperature difference across the material. In practical, the thermoelectric based heat-to-electricity convertion is realized by thermoelectric modules, in which multiple p- and n-type of legs are connected electrically in series and thermally in paralell to increase the voltage and power. The smallest form of a thermoelectric generator (TEG) consists of two electrodes and a thermoelectric material in the middle. Since the generated output power is related to the internal resistance of the device and the achievable temperature gradient, both the electrical and thermal conductivities of the material play a key role in addtion to Seebeck coefficient. Hence, the efficiency of a classic thermoelectric material operated in a TEG is evaluated by a dimensionless figure of merit at the temperature $T$, $ZT = \alpha^2 \sigma T/\kappa$, comprising the key material parameters: Seebeck coefficient $\alpha$, electrical conductivity $\sigma$ and thermal conductivity $\kappa$.

Today, the majority of research on thermoelectrics has strived to optimize the conflicting trends between the three material properties to maximize $ZT$. However, for temperature ranges below 250 °C, the best TEGs are based on $Bi_2Te_3$ ($ZT$=1.2 at 300 K) or similar metal alloys [4]. Such compositions of rare elements are not environmentally friendly, not abundant and thus expensive. Making up half the total cost to build TEGs, the expense of materials prevents TEGs from scaling up to large areas for practical applications. Moreover, the thermoelectric properties of semiconductors have reached their limit from a material perspective [5]. Research is now focused on improving their performance by nanostructuring existing materials, and this represents a substantial production cost [6]. On the other hand, the TEGs are considered promising solution to power wearable sensors and devices [7] over heat engines, because they do not require any moving part and can be miniaturized without loosing the conversion efficiency. The mechanical rigidity of inorganic semiconductor and half-metal alloys limits their application in those portable and wearable devices.

In recent studies, electronic conducting polymers have attracted increasing attention because they are based on abundant elements (C, H, O, S, N) and thus reduce the material cost [8,9]. Despite the tremendous efforts and accomplishments in recent years, the Seebeck coefficient of p-type organic thermoelectric materials at their optimized $ZT$ remains quite low (maximum 70 µV K$^{-1}$ for easily produced polymers, e.g. PEDOT with $\sigma$ = 900 S cm$^{-1}$) [9]; while the ratio between the electrical and thermal conductivities is already similar to Bi$_2$Te$_3$ alloys [10]. The Seebeck coefficient of n-type organic conductors is a bit superior to their p-type conterpart for same conductivity, but their electrical conductivity (only a few S cm$^{-1}$) [11] and air stability still need to be improved. In order to achieve sufficient output voltage, many legs must be fabricated and connected to construct a thermoelectric module. Hence, the manufacturing cost (the other half of the total cost) of conductive polymer based TEG is difficult to reduce. The other challenge of applying organic conductors in thermoelectrics is that their optimum properties are difficult to keep for thick layers, which are necessary for vertical legs to maintain a decent temperature gradient. Hence, organic thermoelectrics are up to now mosty considered for thin film applications, where the device geometry is not optimum for power generation because of the heat leakage through the substrate. Therefore, there is a strong demand for new classes of materials and novel concepts that can help to overcome these current shortcomings.

Similar to the electronic thermoelectric effect, the thermodiffusion of cations and anions in electrolytes can also induce a potential difference along the temprature gradient. The ionic Seebeck coefficient $\alpha_i$ is defined as the ratio between the open circuit voltage and the temperature difference across an electrolyte. Previously, the studies of ionic thermoelectric effect were mainly focused on aqueous solutions of salt, and the values of $\alpha_i$ remained in the range of 0.1–1 mV K$^{-1}$. Nowadays, more and more types of electrolytes with high Seebeck coefficients have been reported, ranging from a few mV K$^{-1}$ up to more than 20 mV K$^{-1}$, which is more than two orders of magnitude larger than that of electronic materials. Of course the high ionic Seebeck coefficient of electrolyte compared to the electronic Seebeck coefficient of organic (semi)conductors is relevant when talking about a circuit comprising a thermoelectric generator in series with an energy storage device (supercapacitor or battery); but it is not enough when talking about thermal-to-electrical conversion efficiency and thermogenerated electrical power.

In this review, we will start with a discussion of the Soret effect, which is the driving force for the ionic thermoelectric effect, followed by the state of art about ionic thermoelectric materials. In the last section, the applications for ionic thermoelectrics are summarized and discussed. The purpose of this review is to introduce ionic thermoelectrics, to present recent progresses in material developments and device innovations, to offer new thoughts for further development of this young research field. We hope this review will inspire not only new designs of ionic thermoelectric materials, but also bringing together scientists with in depth expertise in thermodiffusion phenomena and with scientist experts in electochemical energy storage, bio-signal processing and logic electronics. Together, we will advance our understanding of ion transport and realize applications impossible to reach with classic electronic materials.

## 2. The driving force of the ionic thermoelectric effect
### 2.1. The Soret effect

Applying a temperature gradient to a fluid consisting of several atomic or molecular species, results in a steady state with non-uniform composition. This effect was discovered in the 19$^{th}$ century by Ludwig [12] and Soret [13], who observed that in a tube filled with an electrolyte solution, the salt concentration $c$ was higher at the cold side. Soret realized that this accumulation of salt at lower temperature could not be explained in the framework of Fick diffusion, but required a current component proportional to the temperature gradient. From the steady-condition of zero current, he then obtained the concentration gradient

$$\nabla c = -cS_T \nabla T, \tag{1}$$

where the coefficient $S_T$, now named after Soret, may take either positive or negative sign. Most solutes are "thermophobic" and move towards the cold ($S_T > 0$). The Soret effect was extensively studied historically in view of applications such as isotope separation [14], nuclear fusion in ionized gas [15], optical fiber manufacturing in vacuum deposition processes [16], and separation of the components of liquid hydrocarbon mixtures [17–19]. Several review articles provide an overview of experimental findings and techniques [20–23]. In the last two decades, thermodiffusion has emerged as an versatile contact-free tool at the micro- and nanoscale [24,25]. Coupled to thermal convection, thermodiffusion provides a highly effective trap for DNA [26] and a possible setting for the onset of probiotic chemical reactions [27], whereas laser-heated metal nanostructures are used in opto-thermoeletric tweezers [28] and the generation of micro-flows [29].

### 2.2. The concept of heat of transport

In 1926 Eastman [30] introduced the "heat of transport" $Q^*$, accounting for the change of heat related to the transfer of a solute particle between two regions at slightly different temperature. In this picture, the "heat of transport" $Q^*$ is related to the Soret coefficient according to:

$$S_T = \frac{Q^*}{k_B T^2}. \tag{2}$$

In general, the Soret effect is a non-equilibrium phenomenon involves dissipation which requires the consideration of the underlying transport processes. Onsager's theory for irreversible processes [31,32], published in 1931, provides a formal framework in terms of competing transport phenomena. In the simplest case, the current of the solute molecular species $i$ consists of two contributions:

$$J_i = -L_{ii} k_B T \frac{\nabla c_i}{c_i} - L_{iq} \frac{\nabla T}{T^2}, \tag{3}$$

where the first term describes Fick diffusion with coefficient $L_{ii}$, and the second one thermodiffusion with the Onsager cross coefficient $L_{iq}$ [33]. It turns out convenient to rewrite the former as $L_{ii} = c_i \mu_i$, where $\mu_i$ is the usual mobility as defined through the response to an external force, as for example through the diffusion coefficient $D_i = k_B T \mu_i$ or the sedimentation velocity $\mu_i m_i g$ of a mass $m_i$. For the sake of consistency with Eq. (2), the thermodiffusion coefficient is rewritten as $L_{iq} = c_i\mu_i Q_i^* T$, resulting in
$$J_i = -\mu_i k_B T \nabla c_i - c_i \mu_i Q_i^* \nabla T/T. \tag{4}$$
With the steady-state condition $J_i = 0$ one readily recovers the Eq. (1) with the Soret coefficient from Eq. (2).

The fundamental principle of non-equilibrium thermodynamics is the second law, which states that heat flows to the cold in Clausius's formulation. This is illustrated by rewriting the thermodiffusion current in Eq. (4) as $-c_i v_i$, with the thermal drift velocity $v_i = \mu_i f_i$, arising from the thermodynamic force $f_i = Q_i^* \nabla T/T$; in other words, the heat of transport drags the solute molecule to lower temperature. This picture is based on the assumption that thermodiffusion obeys the same physical mechanism as transport driven by external forces, for example in terms of the Stokes-Einstein friction coefficient $\mu_i = 1/6\pi\eta a_i$ for a spherical particle of radius $a_i$ in a solvent of viscosity $\eta$. The validity of this assumption is discussed in Section 2.4 below.

## 2.3. The ionic thermoelectric effects

When the particles are ions in an electrolyte solution, several novel aspects arise from these mobile charges. Soret's measurements show an accumulation of salt at the cold boundary of a closed vessel, which is described by Eq. (2) with the mean heat of transport of cations and anions ($Q^* = \frac{1}{2}(Q_+^* + Q_-^*)$) (For the sake of simplicity we consider monovalent ions; the relations are readily generalized [34]). In general, the heat of transport of ions differs from each other, $Q_+^* \neq Q_-^*$, and so do the drift velocities $v_\pm$, such that their difference results in a net electric current $I_T = e(c_+ v_+ - c_- v_-)$. In 1889, Nernst discussed transport in "chains of liquids" [35] and showed that an electromotive force and an electric current arise between two identical electrolyte solutions at different temperatures. A schematic view of an open system sandwiched between two non-isothermal reservoirs is presented in Fig. 1(a). Expliciting the thermodiffusion velocities in $I_T$, one obtains the thermally driven electric current through the system:
$$I_T = -\alpha\sigma\nabla T, \tag{5}$$
where $\sigma$ is the ionic conductivity and the coefficient $\alpha$ is given by the difference of the heat of transport between the cation and anion:
$$\alpha = \frac{t_+ Q_+^* - t_- Q_-^*}{Te}, \tag{6}$$
weighted with the Hittorf transport numbers $t_\pm = c_\pm \mu_\pm/(c_+\mu_+ + c_-\mu_-)$. In analogy to thermoelectric effects in electronic conductors, $\alpha$ is addressed as ionic Seebeck coefficient.

In recent years, the thermoelectric effect has been shown to play an important role in colloidal thermophoresis in electrolyte solutions. The heat of transport of the salt ions determines to a large extent whether the colloidal solute diffuses to the hot or to the cold side [36], as observed experimentally for micron-size polystyrene particles [37,38], nano-size micelles [39], and DNA [40]. Yet most experimental work are done in closed systems, as illustrated in Fig. 1(b), where the ion currents vanish and where $I_T = 0$. As a consequence, their thermoelectric properties are not described by Eqs. (5) and (7), but by the thermoelectric field

$$E = \hat{\alpha}\nabla T \qquad (7)$$

with a modified Seebeck coefficient [34]

$$\hat{\alpha} = \frac{c_+ Q_+^* - c_- Q_-^*}{(c+c)Te}, \qquad (8)$$

where the Hittorf numbers $t_\pm$ are replaced with $c_\pm/(c_+ + c_-)$. The different weight factors in $\alpha$ and $\hat{\alpha}$ are related to the ion gradients (see Eq. (1)), which are present in closed systems. Considering the ionic mobilities of common electrolytes, the coefficients $\alpha$ and $\hat{\alpha}$ may differ significantly, and even take opposite signs [34]; they are identical when $\mu_+ = \mu_-$.

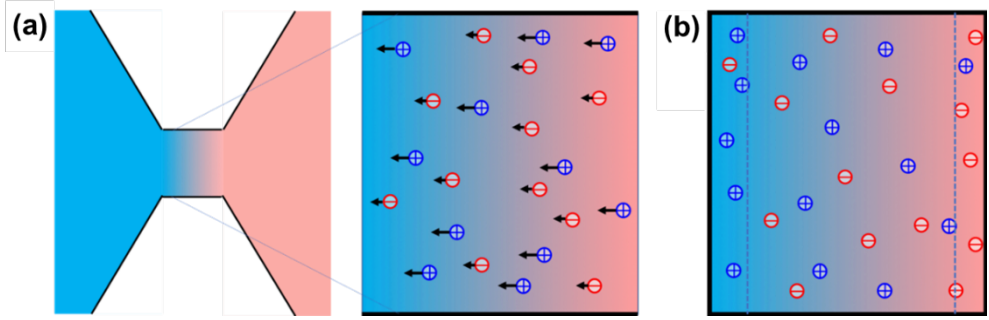

**Fig. 1.** (a) Open system sandwiched between two identical reservoirs, albeit at different temperatures. Both cations and anions diffuse towards the cold, yet the former move faster ($\alpha > 0$) and thus induce an electrical current toward the colder reservoir. (b) Steady-state of a closed system. Positive and negative charges accumulate at the cold and hot boundaries and induce a thermoelectric field (Eq. (7)), such that the currents due to thermodiffusion, gradient diffusion and electric field cancel each other. The coefficient $\hat{\alpha}$ is independent of the ion mobilities, contrary to $\alpha$ in Eq. (6).

Thus in a bulk electrolyte solution subject to non-uniform temperature, there is a macroscopic thermoelectric field described by Eq. (7). This field originates from the thermodiffusion currents $ec_\pm v_\pm$ which permanently transport ions toward lower temperature, deplete and accumulate them at the hot and cold boundaries, respectively. The huge electrostatic interactions forbid charge separation at a macroscopic scale, such that the bulk charge density vanishes, $q(c_+ - c_-) = 0$ and concentrations of positive and negative ions are equal, $c_+ = c_-$. Yet there are net opposite surface charges $\pm\varepsilon\hat{\alpha}\nabla T$ at the hot and cold boundaries within one Debye screening length $\lambda$ ($\varepsilon$ represents the permittivity of the electrolyte) [34]. The coefficient $\hat{\alpha}$ can be determined from the steady-state thermal voltage $\Delta V = -\hat{\alpha}\Delta T$ generated in an ionic thermal cell, which is commonly used as the ionic Seebeck coefficient ($\alpha_i$) in practical device characterizations. The $\alpha$ in open system can be obtained from the transient behavior, where the temperature difference $\Delta T$ is established and the thermal potential isn't yet, the time derivative is given by the relation $d\Delta V/dt = -\alpha\sigma\Delta T/\epsilon$.

It is important to mention that the interface between the electrolyte and electrodes in the closed system disturbs the original ionic distribution. The steady-state thermal voltage obtained in the device results not only from ionic thermodiffusion, but also includes the temperature dependence of the electrode potential [41,42] and specific adsorption [43]. These additional effects depend on the ion species, solvent, and electrode materials; their contributions to the output voltage constitute potential directions for optimizing ionic thermoelectric devices.

*2.4. Explicit theories for the ionic heat of transport in aqueous solution*

Here we discuss the attempts to evaluate the heat of transport of ions in aqueous solution. This overview is restricted to charge interactions, which are much better understood than the contributions due to dispersion forces, hydration, and hydrogen bonds. From the discussion of Eqs. (3) and (4), it is clear that evaluating $Q^*$ in a liquids involves thermodynamic and hydrodynamic aspects. Following Eastman's suggestion $Q^* = TS^*$, Agar et al. identified the heat of transport with the canonical entropy of the solute particle [44], $S = -dG/dT$, with the Gibbs enthalpy $G = \frac{\epsilon}{2} \int dV\, E^2$. With the unscreened electric field of the particle, Agar found:

$$Q^* = \tau \frac{q^2}{8\pi\epsilon a}, \tag{9}$$

where the numerical factor $\tau = -\frac{d\ln\epsilon}{d\ln T}$ accounts for the temperature dependence of the permittivity $\epsilon$. (We use SI units and correct a missing factor 2 in Eq. (4.7) of [44].) More recently, several authors [45–48] confirmed Agar's result, with corrections of the order $a/\lambda$ where $\lambda$ is the Debye screening length. In physical terms, Eq. (9) may be viewed as a polarization energy, resulting from the deformation of the screening cloud in the temperature gradient [48].

In a different approach, one of the present authors started from the Maxwell stress tensor $\mathcal{T}_{kl} = \epsilon E_k E_l (1 - \frac{1}{2}\delta_{kl})$, with the particle's screened electric field $E$ [49,36]. The divergency $\nabla \cdot \mathcal{T}$ gives the force density exerted by the solute on the surrounding liquid, providing the source field for a full hydrodynamic treatment. It turns out that the heat of transport strongly depends on the ratio of particle radius $a$ and screening length $\lambda$. In the limit of small particles it reads

$$Q^* = \left(\tau + (\tau+1)\frac{a}{2\lambda}\right) \frac{q^2}{8\pi a \epsilon}, \qquad (a < \lambda). \tag{10}$$

The first term in parentheses corresponds to Agar's result, with corrections that are not always negligible.

Yet quite a different behavior is found for particles larger than the screening length; simplifying in the weak-coupling approximation one finds [36,49]

$$Q^* = (\tau+1) \frac{\lambda^2}{a^2} \frac{q^2}{8\pi a \epsilon}, \qquad (a > \lambda). \tag{11}$$

As the most striking difference, the heat of transport is reduced by a factor $\lambda^2/a^2$. Noting the size dependence of mobility $\mu \propto 1/a$ and charge $q \propto a^2$, one finds that the thermodiffusion velocity $v = \mu Q^* \nabla/T$ is independent of the particle radius. Eq. (11) compares favorably with experiments on polystyrene microbeads of different size [38]. Several experimental studies confirmed the importance of the additional factor in Eq. (11) and the size-independent thermodiffusion velocity [50]. In electrophoresis, the two limiting cases in Eqs. (10) and (11) are known as Hückel and colloid limits, respectively, and the electrophoretic velocity of large particles is constant with respect to their size, similar to the above result.

Yet the above approaches fail to account for specific-ion effects in the heat of transport of salt ions, which are of the order of at most of few $k_B T$ [44,51–53], yet show large dispersion, e.g., the sodium ion has a value about seven times that of the lithium ion, 3.54 and 0.53 kJ mol$^{-1}$, respectively. This difference indicates that the heat of transport is to a large extent determined by dispersion and hydration forces and, in the case of molecular ions, by hydrogen bonds.

At sufficiently high concentration, ion-ion interactions are comparable to ion-solvent enthalpy, and may significantly affect the heat of transport. Agar and Turner studied the Soret effect of several salts and found that their experimental results for the concentration dependence of the mean heat of transport $Q^* = \frac{1}{2}(Q^*_+ + Q^*_-)$, agreed well with the ionic activities [36]. The variation of the Soret coefficient of nano-scale micelles with concentration [31], was shown to result from their screened electrostatic repulsion [46]. On the other hand, the dependence of the Soret coefficient on the molecular weight of polyelectrolytes [54] and the concentration of silica beads [55], was explained as a collective effect, in terms of the ratio of total charge of salt and colloidal solute [56], which was confirmed later on by experiments on charged nanoparticles [57].

*2.5. The heat of transport of charge carriers with hopping dynamics*

Ionic diffusion in various solid-state matrices and polymer gels shows activated behavior, indicating that the carriers move through jumps between nearby sites, according to $\Gamma = \Gamma_0 e^{-\Delta H/k_B T}$. Very recently one of the present authors [58] proposed a model for the thermoelectric properties, relying on the fact that in a temperature gradient forward and backward jumps between degenerate minima do not occur at the same rate, but result in net thermodiffusion toward the cold. Then the Seebeck coefficient in Eq. (8) is given by the heats of transport $Q^*_\pm = \pm(\Delta H_\pm + k_B T)$, where the activation enthalpies $\Delta H_\pm$ of positive and negative charge carriers may attain values of about 0.5 eV, giving rise to a thermoelectric coefficient of the order of mV K$^{-1}$.

Recently observed values of 10 mV K$^{-1}$, could be related to companion fields induced by the non-uniform temperature, e.g., the concentration gradient $\nabla c = (dc/dT)\nabla T$ if a molecular component such as water. Then the heat of transport acquires an additional term,

$$Q^*_\pm = \pm\left(\Delta H_\pm + k_B T - \frac{d\Delta G_\pm}{dc} T \frac{dc}{dT}\right), \tag{12}$$

with the free enthalpy of activation $\Delta G_\pm$ contains also entropy contributions [58]. Data on the activated hopping dynamics suggest that the correction term my by far exceed the enthalpy of activation and thus result in Seebeck coefficients of 10 mV K$^{-1}$ or more.

In spite of these promising developments, there is at present no satisfactory method to predict the ionic Seebeck coefficient of a given electrolyte. In Section 3, we give an overview of recent experimental findings and review the state of art of ionic thermoelectric materials.

## 3. The state of art of ionic thermoelectric materials

In this section, we will first briefly discuss the characterization of ionic thermoelectric materials, including the Seebeck coefficient, ionic conductivity and thermal conductivity. Afterwards, the recently reported typical ionic thermoelectric materials are presented and discussed.

### 3.1. The characterization of ionic thermoelectric materials

The thermodiffusion behavior of the ions under temperature difference is the most concerned property for an ionic thermoelectric material. Although there are modern techniques that have been successfully applied to study Soret effect of particles and polymers, such as light scattering [59], beam-deflection [24] and fluorescent labelling [60], these methods are not suitable to locate the distribution of small ions. Impedance and other spectroscopy methods could offer some information of the ionic concentration, but they are limited to relatively large-scale devices [61,62]. So far, most of the experimental work is focusing on characterizing the steady-state ionic Seebeck coefficient of electrolytes from the linear fitting of the stable obtained thermal voltage versus the applied temperature difference. Both vertical and lateral structures have been used, depends on the physical properties of the electrolytes. Fig. 2(a) illustrates the vertical characterization setup, in which the electrolyte is sandwiched between two electrodes. Heater and coolers (typically Peltier elements) are in contact with the substrates (or electrodes directly) to control the temperature. The open circuit thermal voltage ($\Delta V_T$) and temperature difference ($\Delta T$) are measured between the two electrodes. This structure is commonly used to characterize and operate devices with liquid electrolytes [63,64]. As shown in Fig. 2(b), in the lateral measurement setup, the studied electrolyte is placed on top of two electrodes exposed to different temperature. Compared to the vertical structure, the lateral one is easier to prepare and requires smaller amount of electrolytes. However, for thin film electrolytes that are open to environment, there can be very large disparity of the water content at the hot side and cold side due to evaporation or hydrophilicity change. Hence, the induced different concentration of mobile ions could greatly affect the measured Seebeck coefficient. In principle, this issue could be alleviated by encapsulating the films during the measurement.

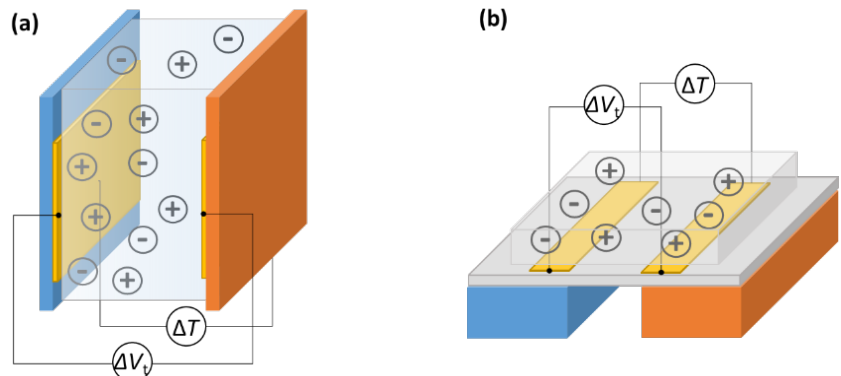

**Fig. 2.** The illustration of the ionic Seebeck coefficient measurement setups. (a) The vertical structure and (b) the lateral structure.

The ionic conductivity is usually measured by using frequency dependent impedance spectroscopy [65]. The mobility of ions can be determined from the impedance in the high frequency range where the phase angle is close to zero, i.e. the system is mainly resistive, and further supported from suitable equivalent circuit fitting if the resistive regime is overlapping with other relaxation phenomena [65]. The thermal conductivity of ionic materials can be measured through typical reported static or transient state methods, such as differential 3ω-method, transient heating method, AC-calorimetry. The research on thermal properties has been relatively well established, existing reviews ([66] as an example) provide the complete view of the field. For composite amorphous samples where the scattering-dominated average phonon wavelength is small, the total thermal conductivity can be obtained from the individual thermal conductivity multiplied with their volume fraction (so called effective medium theory) [67,68]. As a first approximation, the composites are assumed as micro-domains that are large enough to form parallel paths for the heat transport. Such simple approach was used to illustrate the effect of humidity on the thermal conductivity of polyelectrolytes [67].

*3.2. Polyelectrolytes*

The most intuitive way to generate a difference in thermodiffusion between anion and cation is to immobilize one of them. Polyelectrolytes have one type of charge (negative or positive) attached covalently to the polymer chain and are more immobile than the small charge carriers (cation or anion). In symmetric electrolytes containing monovalent salt as shown in Fig. 3(a), both cation and anion can thermodiffuse similarly along the applied temperature gradient. A concentration gradient of the salt will be induced as shown in Fig. 3(c), but no thermal voltage since the ions neutralize each other. In polyelectrolytes (polyanion is used as an example in Fig. 3(b)), the ions that are connected to the backbone of the polymer cannot move as easy as the small ion, and the unbalanced concentration difference will generate a potential difference between the cold and hot side (as shown in Fig. 3(d)).

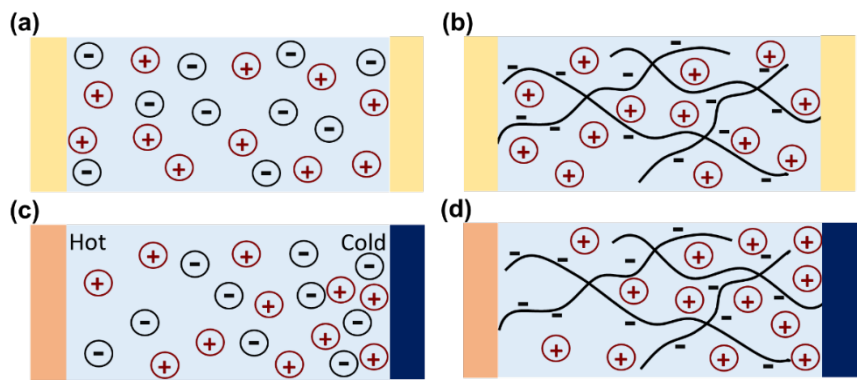

**Fig. 3.** Illustration of ionic thermodiffusion in different electrolytes: (a and c) Electrolytes containing simple monovalent salt, (b and d) electrolytes containing polyanion and small counter cation.

**Table 1.** Thermoelectric properties of some polyelectrolytes.

| Material | $\alpha_i$ at low/high RH[a] (mV) | $\sigma_i$ at low/high RH (S m$^{-1}$) | Power factor ($\mu$W m$^{-1+}$ K$^{-1}$) | Year |
|---|---|---|---|---|
| Nafion-Ag [69] | 0.3/1.1 mV | 0.002/0.05 | 0.06 | 2015 |
| PSS-Ag [69] | −0.1/−2.1 | 0.01/0.4 | 1.76 | 2015 |
| PSSH [70] | 5.1/7.9 | -/9 S | 562 | 2016 |
| PSSNa [71] | 0.3/4.1 | 0.02/0.9 | 15 | 2017 |
| Nafion [72] | 4.2/3.6 | 0.3/1.9 | 24 | 2018 |
| S-PEEK [72] | 1.2/5.5 | 0.008/0.2 | 6 | 2018 |
| PDDAc [72] | 10/18 | 0.01/1.9 | 615 | 2018 |

[a] low humidity is of the range of RH= 30% to 50% and high humidity is of RH=70%.

Table 1 compares the Seebeck coefficient and ionic conductivity of some typical polyelectrolytes that have been reported recently. These polyelectrolytes are prepared as solid states film; however, they can absorb small amount of water from the air due to hygroscopic effect. Hence, the ionic transport mechanism of these polyelectrolytes is affected by the formation of percolation path, and both the conductivity and Seebeck coefficient are humidity dependent. In 2015, Chang et al. reported the thermoelectric properties of single ion conductors composed of silver Nafion (Nafion-Ag) and silver polystyrene-sulfonate (PSS-Ag) [69]. The ionic transport under a temperature gradient was discussed on base of the entropy of the chemical reaction and the Soret effect. One of the surprising finding in the work is that the sign of the Seebeck coefficient of Ag-Nafion changed to negative at high humidity (PSS-Ag has positive Seebeck coefficient in the whole studied humidity range), even only cations in the electrolyte are expected to be mobile. The authors explained that the more readily thermodiffusion of water than Ag$^+$ in percolated channels in Nafion resulted in silver flux from cold side to hot side, which is perceived as a negative Seebeck coefficient. Despite the small value of the power factor, this work was the first to report the thermoelectric effect in polyelectrolytes, and discovered the effect of water to the ionic thermal voltage. Later on, more types of polyelectrolyte based ionic thermoelectric materials were reported with enhanced performance and deepened understanding [70–72]. As shown in Table 1, the Seebeck coefficient of these materials is in the order of a few millivolt, and the ionic conductivity spreads over a wide range. For polyanions, the proton conductors have relatively high conductivity than other cations due to the small size of the ions and the fast Grotthuss mechanism in humid environment. By studying the Seebeck coefficient of different types of polyelectrolytes, and investigating the water contact angle of those polymers at different temperature, Kim et al. further analyzed the contribution of the water diffusion [72]. They claimed that at low humidity level (Fig. 4(a)), the transport of water is trivial due to the lack of percolation path. With increasing humidity, the diffusion of water in the electrolyte associates with the solvated ion and generates a potential in addition to the thermodiffusion of ion. Such additional contribution from the Soret effect of water depends on the chemical nature of the nanoporosity; thus could be in the opposite direction to the ionic thermodiffusion, as Nafion in Fig. 4(b), or in the same direction as shown for S-PEEK in Fig. 4(c). They also discovered that the different water uptaking ability at the hot and cold side of the (poly(diallyldimethylammonium chloride) (PDDAc)) could lead to different Cl$^-$ dissociation, which is the main contribution of the thermal voltage in this polycation electrolyte [72], instead of the thermodiffusion of Cl⁻.

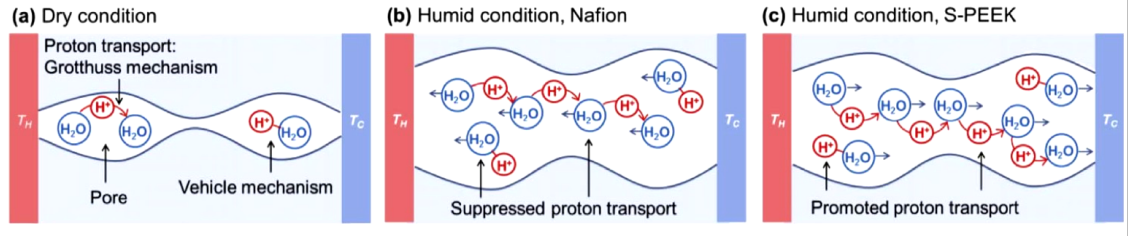

**Fig. 4.** Schematic illustration of ion and water transport in different electrolytes under a temperature gradient at different humidity level [72]. (a) Under dry condition, water transport is negligible due to the poorly percolated water channels. Under humid condition, (b) the thermo-diffusion of the water and ions are in the opposite direction (Nafion as the example), and (c) in the same direction (S-PEEK as the example). (Reproduced with permission from Ref. [72], © Elsevier 2016.)

The water processability of polyelectrolyte materials is convenient for manufacturing with regard of environment and safety issues; however, it also brings main drawbacks. First, the output thermal voltage is not only proportional to the temperature difference, but also highly depends on humidity value, which limits the thermal sensing applications of the materials. Second, since the ion transport largely relies on the percolation channel that requires certain amount of water, these materials are only suitable for thermoelectric applications close to room temperature. The evaporation of water in long time operation at higher temperature will greatly undermine the performance.

*3.3. Polyelectrolytes in conducting polymers*

The ionic transport and its coupling with electronic transport in conjugated polymers are well known [73–75]. The ionic thermoelectric effect in such mixed conductors was first investigated by Wang et al. in 2015 [76]. As shown in Fig. 5(a), the Seebeck coefficient of five different poly(3,4-ethylenedioxythiophene) (PEDOT) derivatives was investigated at different humidity level. For the ionic-electronic mixed conductors, large increase in the Seebeck coefficient at high humidity levels was observed, which is identified as an ionic Seebeck effect. They found that the ionic Seebeck effect in polymeric mixed conductors PEDOT:PSS enhanced the maximum power factor by two to four orders of magnitudes. However, the benefit is not stable over time since the output power dramatically drops (as shown in Fig. 5(b)). The authors proposed that this effect could be used advantageously to improve the heat to electricity conversion efficiency for intermittent heat sources. Ail et al. further revealed the mechanism of the thermodiffusion of charge carriers in PEDOT:PSS mix conductor [77]. As shown in Fig. 6, they investigated the time dependence of the thermal voltage in pure ionic, pure electronic and mixed conductors, and identified the presence of a complex reorganization of the charge carriers. In-situ infrared spectroscopy was used to visualize the thermodiffusion of protons after the temperature gradient was applied, followed by an internal electrochemical reaction.

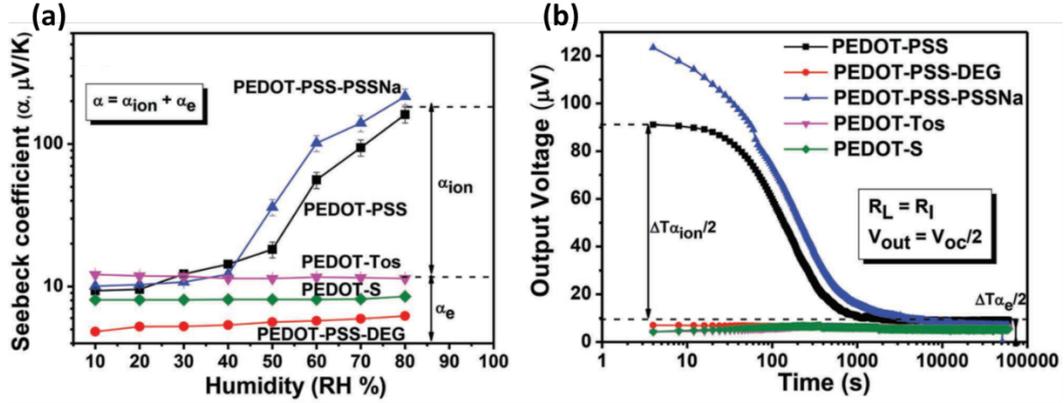

**Fig. 5.** (a) Seebeck coefficient (*α*) of PEDOT-Tos, PEDOT-PSS-DEG, PEDOT-S, PEDOT-PSS, and PEDOT-PSS-PSSNa at different humidity (measured between two gold electrodes). (b) The output voltage over load (equivalent to the internal resistance of the device) for the single leg thermoelectric generators using various materials at 80% RH [76]. (Reproduced with permission from Ref. [76], © Wiley-VCH 2015.)

In order to extend the time scale of the contribution from the large ionic Seebeck effect, different strategies have been proposed. In 2016, Chang et al. used redox ion pairs that can be reduced and oxidized at the electrodes in the mixed PEDOT:PSS conductors to control the duration of the ionic enhancement [78]. They demonstrated that a well-designed material can realized long-lived ionic Seebeck enhancements. Jiang et al. have demonstrated that ionic thermodiffusion could facilitate to generate a large quasi-constant power in hydrazine doped perylene bisimide [79]. The key is that the diffusion of ions in such mixed conductors can be considered as quasi-frozen, because it is much slower than the electrons and holes in the crystalline structures. It is worth to mention that this work showed that efficient n-type of thermoelectric materials can be designed by the rational control of the mobility of the electronic and ionic carriers. Moreover, Guan et al. coated the highly conductive PEDOT:PSS film with PSSH, and enhanced the power factor by several orders of magnitude compared to PEDOT:PSS mixed with PSSH [80]. The high conductivity obtained by $H_2SO_4$ pretreatment was not affected by the additional PSSH layer on top, while the ionic thermodiffusion in PSSH could enhance the thermal voltage in PEDOT ascribed to the energy filtering of the charge carriers of PEDOT:PSS. Similar enhancement of the Seebeck coefficient of PEDOT has been realized with ionic liquid coating [81].

There are more examples of the mixed conductors that have extraordinary good thermoelectric performance. Kim et al. doped PEDOT:PSS film with additional poly(styrene sulfonic acid) (PSSH), and enhanced both ionic Seebeck coefficient and conductivity at high humidity level [82]. In the same work, the authors also aimed to design a device structure to damp the fluctuation in the environmental humidity. A self-humidifying bilayer composed of metal organic framework and hydrogel layer was used to provide stable high humidity (90%) for the thermoelectric device, and a stable output thermal voltage over 72 h in ambient condition proved the feasibility of this strategy. The same group developed the first n-type mixed ionic-electronic composite by doping PEDOT:PSS with $CuCl_2$ [61]. The binding between $Cu^{2+}$ and PEDOT (confirmed by electron spin resonance spectroscopy) facilitated the formation of transport channels in which the thermodiffusion of Cl⁻ resulted in a large thermal voltage. Fluorescence imaging using Cl⁻ as an indicator and time-of-flight secondary ion mass spectrometry mapping confirmed the diffusion of Cl⁻ in the film from the hot to the cold electrode. The n-type mixed conductor shows a high Seebeck coefficient value of −18.2 mV K⁻¹ at 80% humidity level. Aside from PEDOT, mixed conductors composed of polyaniline composited with the anionic polyelectrolyte (poly(2-acrylamido-2-methyl-1- propanesulfonic acid)) and phytic acid were reported in 2020 [82]. Aside from good thermoelectric performance, the proposed materials show remarkable stretchability (up to 750%) and self-healing. The authors demonstrated that the performance of the ionic thermoelectric device was sustainable under 50% strain deformation and 30 cycles of self-healing [83]. This is the first demonstration of such performance in ionic thermoelectric materials, which is promising in wearable applications. The combination of conductive polymers and electrolytes generate various possibilities for charge carrier transport and interactions, which offers great opportunities for the design of ionic thermoelectric materials.

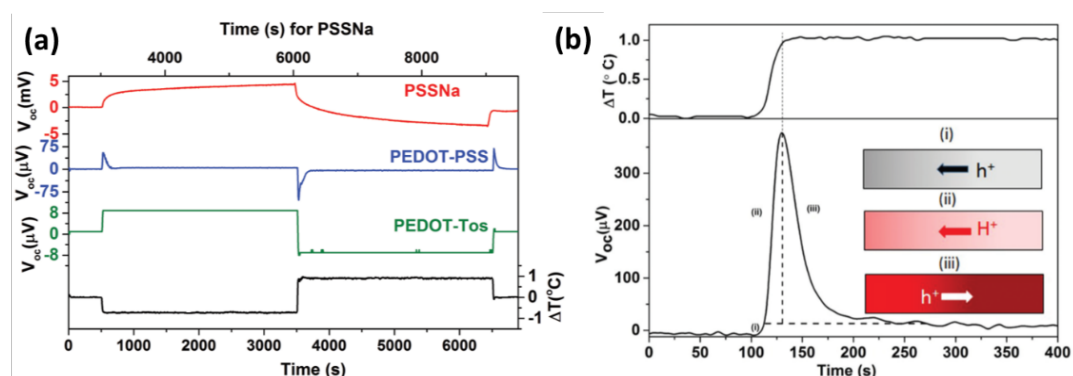

**Fig. 6**. (a) Open-circuit voltage versus time for PEDOT-Tos, PEDOT-PSS, and PSSNa, at 80% RH with Δ$T$ = 1 °C. (b) Different regions of the $V_{oc}$ -time curve under a Δ$T$ and the corresponding mechanism [76]. (Reproduced with permission from Ref. [76], © Wiley-VCH 2016.)

*3.4. Ionic liquid-based electrolytes*

The electrolytes introduced in the previous two sections were water-based materials. The large Seebeck coefficients of the electrolytes composed of ionic liquids in organic solvent up to 7 mV K⁻¹ were reported by Bonetti et al. in 2011 [63]. In contrast to the polyelectrolyte ionic conductors, both the cations and anions are mobile in the ionic liquid solutions. Based on the experimental results and theoretical calculations, the authors pointed out that the large Seebeck coefficient originates mainly from kosmotrope or structure making effects [84]. In 2016, Jia et al. reported their study on the thermoelectric properties of a series of ionic liquids [85]. They theoretically analyzed the ionic desorption from electrode surface and concluded that the different Seebeck coefficients were due to the competition between ion-electrode and cation-anion interactions. The thermoelectric performance of quasi solid state ionic liquid polymer gel was first reported by Zhao et al. in 2019 [68]. As shown in Fig. 7(a), the ionic liquid is soluble in the poly(hexafluoropropylene) (HFP) amorphous phase of the co-polymer, while the other poly(vinylidene fluoride)(PVDF) crystalline phase can maintain the mechanical strength of the gel. The ionic Seebeck coefficient of the ionic gel could be tuned from −4 mV K$^{−1}$ to +14 mV K$^{−1}$ (as shown in Fig. 7(b)) by adding liquid neutral polyethylene glycol. They proved that changing the interaction between ions the polymer matrix led to a complete change in the dominating thermodiffused ionic current from anions to cations. The first printed ionic thermoelectric electric module (or ionic thermopile) based on the p- and n-type of gels in vertical structure demonstrated a thermal voltage of 333 mV K$^{−1}$ for just 18 pairs of legs.

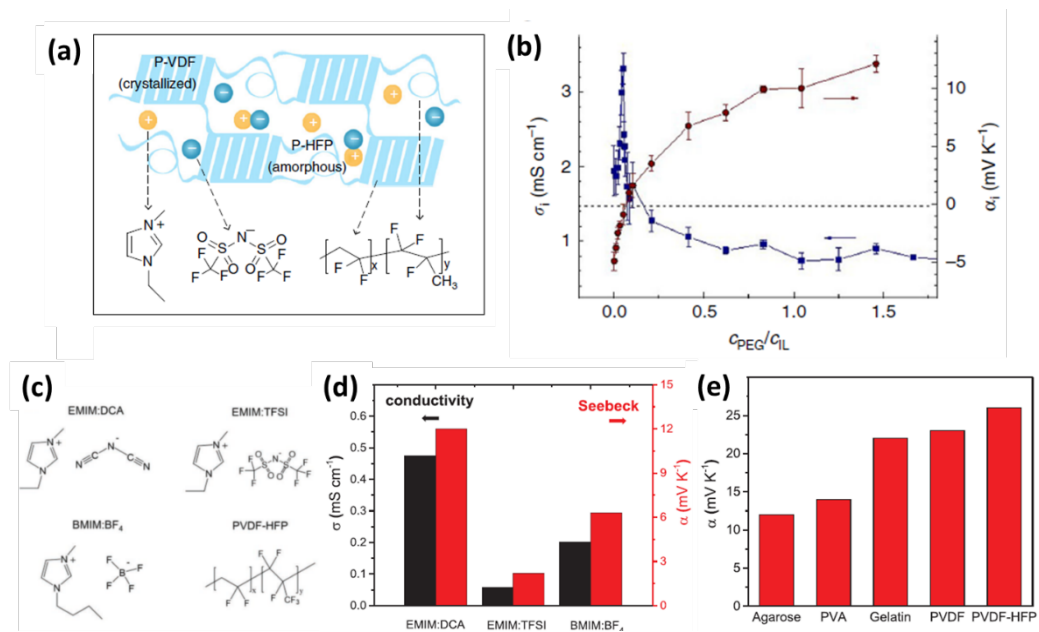

**Fig. 7**. Ionic thermoelectric properties of ionic liquid polymer gels. (a) Composition and (b) the thermoelectric properties of [EMIM][TFSI]/PVDF-HFP polymer gels as a function of the molar concentration ratio cPEG/cIL [68], conductivity is shown in blue squares and Seebeck coefficient in red dots, dashed line indicates Seebeck coefficient = 0 mV K$^{−1}$. (Reproduced with permission from Ref. [68], © Springer Nature 2019.) (c) Chemical structures of ionic liquids, EMIM:DCA, EMIM:TFSI, BMIM:BF$_4$, and PVDF-HFP. (d) The ionic conductivity and Seebeck coefficient of ionogels containing 50 wt% EMIM:DCA, EMIM:TFSI, and BMIM:BF$_4$. (e) Seebeck coefficients of ionogels with the EMIM:DCA loading of 80 wt% with various polymers [86] (Reproduced with permission from Ref. [86], © Wiley-VCH 2019.)

Later on, Cheng et al. reported a few more ionic liquid polymer gels (shown in Fig. 7(c–e)), in which the highest Seebeck coefficient is up to 26.1 mV K$^{−1}$ [86]. The giant Seebeck coefficient was explained to be related to the ion-dipole interaction between fluoride-based polymer and ionic liquids. The incorporation of the ionic liquid in the co-polymer matrix enhances the Seebeck coefficient, but lowers the ionic conductivity of the ILs. In order to maintain the ionic conductivity, He et al. dispersed SiO$_2$ nanoparticles in the ionic liquid to form quasi-solid state ionogels with high Seebeck coefficient of 14.8 mV K$^{−1}$ and obtained a much improved ionic conductivity of 47.5 mS cm$^{-1}$ [87]. The mechanical properties of the gels were also investigated through dynamic mechanical measurements. The remarkable ionic Seebeck coefficient and ionic conductivity are attributed to the interactions between ionic liquid and the surface of the SiO$_2$ nanoparticles.

The key advantage of the water-free ionic liquid-based electrolytes is their stability upon long time operation under heating. The "ambipolar" composition makes it possible to design p- and n-types of materials by controlling the ion-polymer interaction. However, the ionic conductivity of most ionic polymer gels are at least 2–3 orders of magnitude lower than water-based electrolytes, which is main drawback of this type of materials. The thermodiffusion of ions with their kosmotrope and chaotrope effects is far from completely understood in non-aqueous systems [63].

*3.5. Hybrid materials*

As mentioned earlier, the ionic Soret coefficient is sensitive to the interaction of the ion with its environment, surprisingly high Seebeck coefficient can arise in hybrid electrolytes due to additional interactions involving the mobile ions. Zhao et al. used a water-free electrolyte based on NaOH treated low molecule weight polyethylene oxide (PEO) in their thermoelectric devices [64,88]. The complex polymer-based composite was completely different from previous studied aqueous solutions, the heat of transport of Na$^+$ and OH$^-$ could not explain the large positive Seebeck coefficient obtained (~11 mV K$^{-1}$). The authors used the Born model [89] to account for the electrostatic contribution from the solution to the thermal voltage, but together with the calculated Seebeck coefficient obtained from the heat of transport of the mobile ions [90], they were unable to explain the large ionic Seebeck coefficient. Chemical analysis indicated the formation of polyanion from the deprotonation of PEO, which could be the reason for the observed large Seebeck coefficient. However, this is not yet the end of the story. Li and Hu et al. discovered an electrolyte consisting of polyethylene oxide PEO, deionized water and NaOH that shows high Seebeck coefficient (10 mV K$^{-1}$) [91]. With the presence of water, the sodium alkoxide would recover back to hydroxyl group and NaOH, this result weakened the assumption that polyanion was the main reason for the large Seebeck coefficient in NaOH-PEO electrolyte. Aside from the discussion included in the publications, there are other factors might have great impact to the thermodiffusion in those complex systems. For example, previous studies show that the hydrogen bond capability of a solute plays a key role in the thermodiffusion [92].

Furthermore, the Seebeck coefficient of PEO and NaOH in water solution can be enhanced to 24 mV K$^{-1}$ by infiltrating the electrolyte into the cellulosic membrane with nanochannels. The authors compared the thermal voltage response of the same electrolytes filtered in membranes with different charge density, and attributed the enhanced thermal voltage to effective sodium ion insertion into the charged molecular chains of the cellulosic membrane [91]. This work demonstrated the feasibility of using interface effect and nanoporosity to modify the thermodiffusion of ions, which offers a new avenue to develop the fundamental understanding and applications.

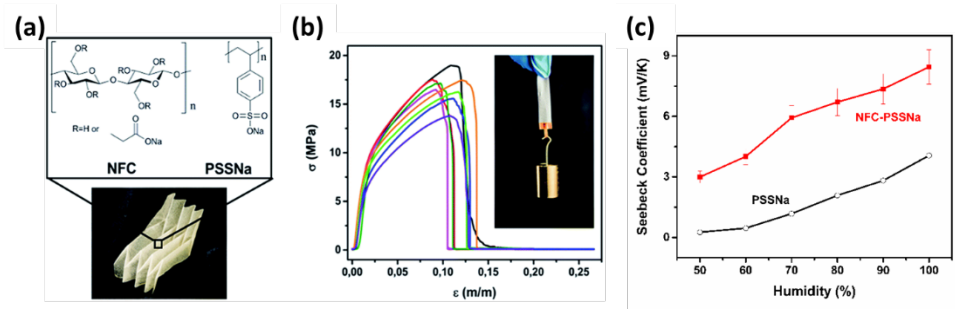

**Fig. 8.** Ionic thermoelectric paper. (a) Photograph of flexible NFC-PSSNa paper and the chemical structure of nanofibrillated cellulose (NFC) and polystyrene sulfonate sodium (PSSNa). (b) Tensile strength versus strain for eight different NFC-PSSNa samples with the same composition and thickness. (c) The Seebeck coefficients of the hybrid material compare to pure PSSNa film at different humidity level [67]. (Reprint from © Royal Society of Chemistry 2017)

The approach of making nanocomposite was used to improve the mechanical properties of the ionic thermoelectric materials and extend their applications. Jiao et al. presented an ionic thermoelectric paper by integrating polyelectrolytes with nanofibrillated cellulose (NFC) [67]. As shown in Fig. 8 (a and b), the resulting NFC-PSSNa paper is flexible and mechanically robust, which is desirable for manufacturing and utilization. Moreover, as shown in Fig. 8(c), the ionic Seebeck coefficient of the robust NFC-PSSNa thermoelectric paper is higher compared to pure PSSNa film, especially in low humidity level. This might due the good moisture retention of the NFC and the formation of water percolation path along NFC fibers.

In the field of thermogalvanic cells (TGCs), hybrid systems have been well applied to enhance the performance and processability, e.g. reduce the leaking risk in practical devices [93]. Recently works showed that introducing additional components could even greatly increase the Seebeck coefficient of the redox ion pairs. Han. et al. discovered that the thermodiffusion of non-redox ions from added salt could modulate the Seebeck coefficient of the redox couple in TGCs [94], and using the optimized salt could lead to a great enhancement of the output power. Yu et al. hybridized TGCs by adding molecules that introduce crystallization of the redox active ions. They demonstrated that the deposition toward the bottom of the cell could facilitate ionic transport and improved the output voltage by 2–3 times [95]. These works show the great potential of hybrid systems that could be one of the promising strategy for future improvement for ionic thermoelectric materials.

## 4. Applications of the ionic thermoelectric effect and challenges

The investigation of ionic thermodiffusion has been driven mostly by scientific curiosity rather the societal needs since no energy devices had been demonstrated based on that effect. However, after the first novel proof of concept that the ionic thermal voltage can be used for energy harvesting, there has been a growth in the number of publications in this new research field. In this section, we discussed the recent progresses in the application of ionic thermoelectrics, energy harvesting and heat/thermal sensing.

*4.1. Thermal energy conversion*

Energy conversion using electronic thermoelectric effects is a well-developed research area with products on the market comprising inorganic materials. For a traditional thermoelectric leg composed of a semiconductor and two metal electrodes exposed in a temperature gradient, a constant electrical power can be generated to an external load through the electrodes. The mechanism is different for ionic thermoelectric materials because cations and anions are the charge carriers instead of electrons and holes in semiconductors. The thermodiffused ions are blocked at the surface of the metal electrode and cannot pass through the external circuit.

There are two concepts that can be used to generate a current from the accumulated ions under temperature gradient. It depends if the ions are redox active or not. The first is thermogalvanic cells [96] in which an electron transfer takes place between the electrode and the redox ions located at the electrode/electrolyte interface. The redox ions act as shuttles for the electrons passing across the electrolyte and being delivered at the other electrode to produce a continuous current. The thermogalvanic effect leads to a true thermoelectric generator with constant current under the temperature gradient. Those devices are found in both aqueous [97,98] and organic solution [99,100] using various inorganic and organic redox couples. We are not going to cover this field of research in this review. Note that the redox reaction can also take place inside electrodes via insertion/deintercalation of ions, e.g., lithium in an intercalated compound such as $Li_xV_2O_5$ [101] and several layered oxide [102] leading to a thermally rechargeable battery. In that case the thermal voltage is generated by redox reaction at the electrode but the current vanishes when the device is charged. The redox electrode material and the electrolyte must match for good interfacial stability; which limits the number of electrolytes.

The second concept utilizes the ionic Seebeck effect with ions that are not redox active; which is the focus of this review. The strategy is to use the developed ionic thermal voltage as internal voltage source in the device and charge the electric double layer capacitors (EDLCs) at electrodes with large capacitance (e.g. with large internal surface areas). In this section, we will review the working mechanism of the ionic thermoelectric charged supercapacitor (ITESC), different operating modes and the figure-of merit of the device.

*4.1.1. Ionic thermoelectric charging mechanism*

In an ionic thermoelectric cell, the thermodiffused ions accumulated at the electrolyte/electrode interface could lead to a current flow in the external circuit. This forms the basic principle of an ionic thermoelectric supercapacitor (ITESC) that realizes energy harvesting from ionic thermoelectric materials [64,71]. The detailed complete operating mechanism of ITESC can be divided into four steps as shown in Fig. 9(a). The step i is the preparation of charging, in which a $\Delta T$ is applied to the device and the corresponding open circuit thermal voltage is generated. In the demonstration device in Fig. 9, a positive thermal voltage is measured at the cold side of the cell due to the dominating thermodiffusion of sodium cation [64]. The step ii is the charging of EDLC by connecting the two electrodes, which can be through a load or in short circuit. During this step, a current passes from the high electric potential electrode to the low potential side while $\Delta T$ is kept. The step iii is the equilibration period, in which the $\Delta T$ and the load are both removed from the circuit. The key point is that the ions diffused back but the charges stored at the electrode-electrolyte interface remain. Since no thermal voltage is produced, the resulting open circuit voltage at $\Delta T=0$ is governed by the stored charge at the electrodes, thus with an opposite sign compared to the original thermal voltage in step i should be seen in this step. The step iv is the discharge of EDLC through a load resistance in the external circuit, i.e. the consumption of the stored charge. The temperature difference is only needed for steps i and ii, and not needed after the charging process finished, which enables energy storage and utilization without the heat source.

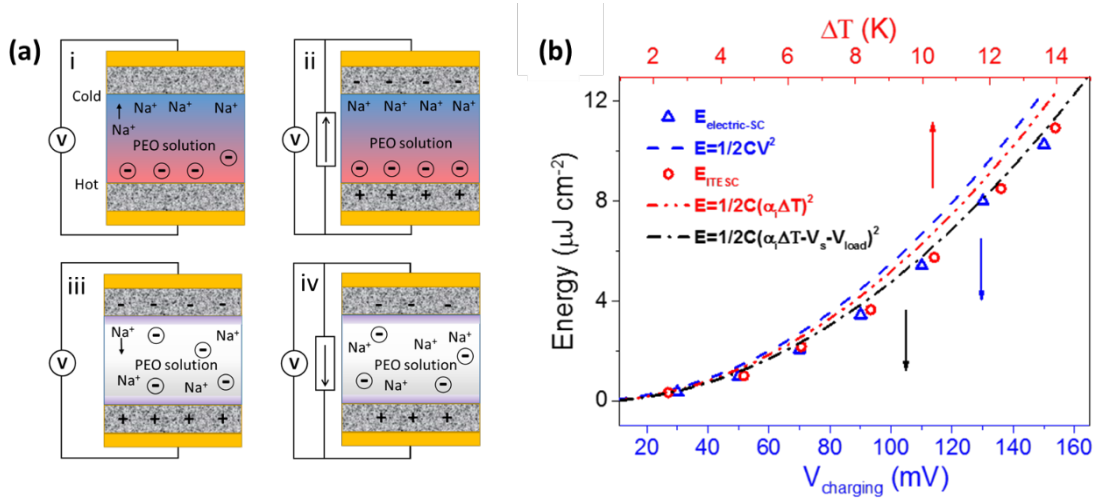

**Fig. 9.** Operation principle of ionic thermoelectric supercapacitor (ITESC). (a) The mechanism sketch (b) of a full charging and discharging cycle: (i) Applying $\Delta T$ to establish the ionic thermal voltage, (ii) thermoelectric charging of the supercapacitor, (iii) removing $\Delta T$ for the ions to reach equilibration and (iv) discharging [64]. (b) Energy density versus $\Delta T$ for the ITESC (red open circles and dashed line), electric charging (blue open squares and dashed line) (Reproduced with permission from Ref. [64], © Royal Society of Chemistry 2016.).

To verify the charging mechanism, the correlation between the ionic Seebeck coefficient (generated thermal voltage) of the electrolyte and the stored electric energy in the supercapacitor was analyzed. An ITSEC was charged and discharged under different $\Delta T$, and the energy flow into the supercapacitor was calculated by integrating the current over a load resistance with time. As shown in Fig. 9(b), the measured stored energy (red dots) increases quadratically with $\Delta T$ (red dashed line):

$$E = \frac{1}{2}C(\Delta T\alpha)^2 \tag{13}$$

The stored energy in the same device charged by external power supply plotted versus charging potential (blue line and triangle) overlapped with the thermally charged curve. This work proves that the generated/stored electric energy increase quadratically with the ionic Seebeck coefficient that directly related to the charging potential via $V = \Delta T \alpha$.

*4.1.2. Different operating modes*

For the ionic thermoelectric supercapacitor (ITESC) [64,71], the typical operation curve corresponding to the presented four steps is shown in Fig. 10a. In the first step of charging preparation, a thermal voltage is generated in open circuit and saturated under $\Delta T$. In the second step, the current passing through the connected load is measured during the charging process. The time needed for the current to decay to 0 depends on the capacitance of the electrodes and the total resistance (internal and load resistance). The integration of the current ($I$) and power ($I^2R$) give the accumulated charge and electric energy during charging, respectively. In the equilibration period in step iii, the measured open circuit voltage between the two electrodes reverses sign compare to step i, since the ions diffused back but the charges stored at the electrode-electrolyte interface remain. In an ideal ITESC without self-discharging, the reversed open voltage should reach identical value as in step i. Hence, the discharging current in the last step through the external circuit will be the same as the charging current in step ii, i.e. the consumption of the stored charge. After the complete cycle, the ITESC returns to its original state.

The first device demonstrating this concept was not perfect; and a significant self-discharge was observed [64]. This was improved quickly in the following work, in which the thermally induced voltage in PSSH film enabled electrochemical reactions in polyaniline-coated electrodes containing graphene and carbon nanotube [70]. The thermodiffused protons can oxidize or reduce the polyaniline film as a pseudo-supercapacitor, and the areal capacitance of 120 mF cm$^{-2}$ was achieved. The author showed that the charge stored in such supercapacitor was stable for over 24 hours without obvious self-discharging. Same device principle can be operated with load resistance for the whole cycles, in which symmetric charging and discharging current can be generated through the load when $\Delta T$ was applied periodically [86]. This work successfully proved that ITESC can be used to collect thermal energy and store as electricity.

Another reported concept considers the thermodiffusion of ions in the electrolyte as the process of charging the EDLC is illustrated in Fig. 9(b) [103]. The charging process in this type of devices is when the ions diffuse under a temperature difference and establishes the thermal voltage in open circuit. To discharge the device, an external load is connected between the cold and hot electrodes, which enables a current flow from the high potential to low potential. An example of the charging and discharging curve in such device is shown in Fig. 9(b), the temperature difference was applied during the whole charging and discharging period [103]. In later work using electrolyte filled cellulose membrane, similar converting principle was used to demonstrate the device [91]. The charging process was presented as the development of the thermal voltage in electrolytes, and the discharging process was conducted by applying constant current of 500 nA cm$^{-2}$ while keeping the temperature difference. This type of device was claimed to generate electricity from constant heating source, however, without resetting of the thermodiffused ions, the cyclability of the device can only be realized with quick self-discharging. Moreover, it cannot store electricity because the discharging step is not separated from the charging, equivalent to a capacitor that can only be used while charging, which limit its applications.

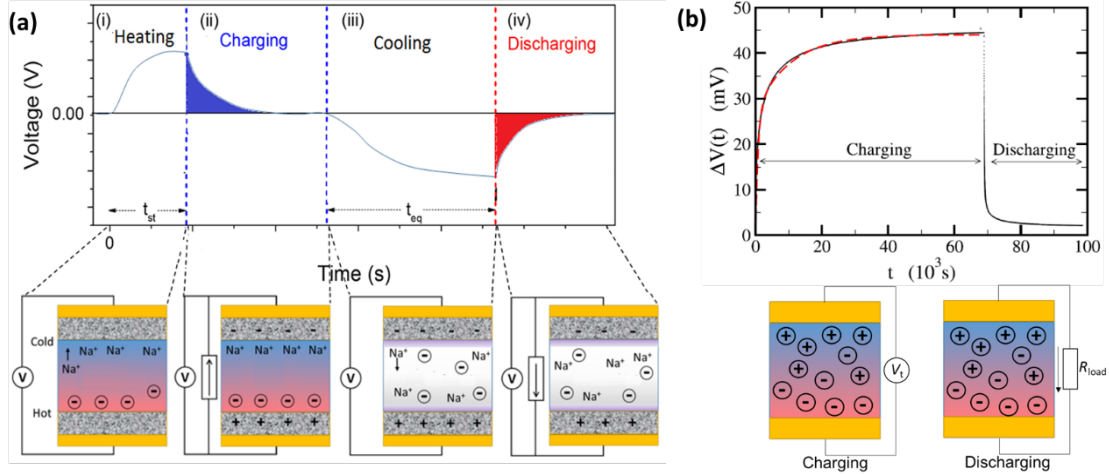

**Fig. 10.** Different operation modes of ionic thermoelectric charging device. (a) Charging with $\Delta T$ and discharge at $\Delta T=0$ [70] (Reproduced with permission from Ref. [70], © Wiley-VCH 2017). (b) Charging and discharging both with $\Delta T$ [103]. (Reproduced with permission from Ref. [103], © AIP Publishing 2015.)

*4.1.3. Ionic thermoelectric figure of merit*

The figure of merit $ZT_i = \sigma_i \alpha_i^2 T / \lambda$ has been used to describe the efficiency of ionic thermoelectric materials in analogy with electronic thermoelectric materials. However, The $ZT_i$ for electrolytes is not related to the heat-to-electron conversion efficiency the same way as the classic thermoelectric materials, since ionic materials cannot be applied in traditional TEGs. In 2017, Zhao et al. reported a detailed study on the figure of merit of ITSEC [71]. The charging efficiency was derived from the ratio between stored electric energy and the absorbed heat during charging, and the final expression is shown in Eq. (14).

$$\eta_{ch} = \frac{\Delta T}{T_H} \frac{ZT_i}{2ZT_i + \frac{10T}{T_H} - \frac{1}{2}ZT_i \frac{\Delta T}{T_H}} \tag{14}$$

Using this equation, the $ZT_i$ of an electrolyte is directly connected to the efficiency of an ITESC composed of this material. Ionic and electronic thermoelectric materials cannot be compared directly, because the mode of operation of a TEG and an ITESC are different. However, ionic materials in an ITESC of capacitance $C$ can be compared to electronic material in a TEG coupled in series to a supercapacitor (SC) of capacitance C because they have similar equivalent circuits, as shown in Fig. 11(a). Hence, the energy conversion efficiency of ionic and electronic thermoelectric materials can be presented together in the same graph through the equivalence of an ITESCs with a series circuit TEG-SC. As shown in Fig. 11(b), the heat-to-stored electricity conversion efficiency of an ITESC using electrolytes is still lower than $Bi_2Te_3$ in a TEG-SC circuit because of the orders of magnitude lower ionic conductivity. On the other hand, as shown in Fig. 11(c), electrolytes enable the storage of many orders of magnitude larger electrical energy per leg/per $\Delta T$. This is because the stored energy is related to the square of the Seebeck coefficient (see Eq. (13)), which is much larger for polymer electrolytes (~10000 µV K$^{-1}$) than Bi$_2$Te$_3$ alloys (~200 µV K$^{-1}$). In the future material development, it is important to improve the ionic conductivity while achieving high ionic Seebeck coefficient. Note that serial circuit based on TEG charging a SC, or the single ITESC device, are able to provide pulse of electrical power through the discharge of the SC and circumvent the small constant power generated by the TEG.

Aside from the material improvement, to realize the energy conversion of ionic thermoelectrics for practical application, there are more challenges regarding the device design. One practical issue is that they cannot generate continues output power due to the requirement of discharging without temperature difference. Ionic thermoelectric concepts are more suitable to convert intermittent heating/cooling. With suitable design, the charging time for the supercapacitor could match up with the periodic length of the heat source. It is possible to combine ionic thermoelectric devices with other constant power conversion facilities, such as TEGs and solar cells, which cannot function without heating/illumination. The low energy density of ionic thermoelectric devices is another concern, which can be enhanced by manipulating the circuit, e.g. used the ITESC to charge other capacitors, and change the connection from parallel to series when utilizing the energy.

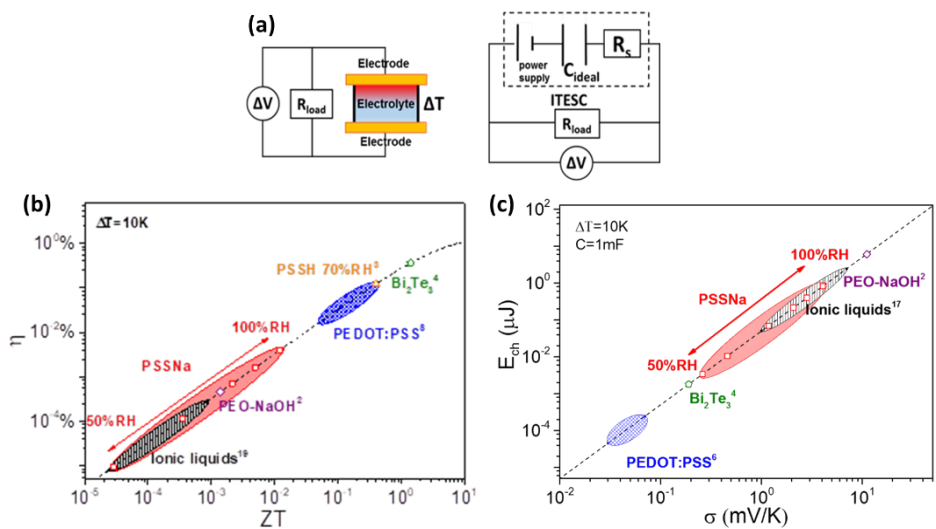

**Fig. 11.** The comparison of ionic and electronic thermoelectric materials in ITESC. (a) Measurement set-up for ITESC (left) and the equivalent circuit of ITESC in the experimental set-up (right). (b) The efficiency and (c) stored energy in the same supercapacitor of materials with different *ZT* [71]. (Reproduced with permission from Ref. [71], © Wiley-VCH 2017.)

Other types of thermal energy harvesting concepts based on thermally induced capacitive effect between the same electrodes at different temperature (no heat flow) [104–108] or asymmetric electrodes [109,110] have been reported by a few groups. Similar to ionic thermoelectric devices, these devices are mostly composed of cost-effective materials and functional for low-grade heat conversion. Actually, the effect of electrode polarization at different temperature is included in the measured thermal voltage; however, further optimization of the electrode might enhance this contribution to the total output voltage.

*4.2. Temperature/heat sensing*

The electronic Seebeck effect has been applied in temperature sensing applications since ages in the so-called thermocouple (TC). The principle of TCs is based on the difference in Seebeck coefficients between two electronic conductors forming a junction, typically one is chosen with a positive Seebeck coefficient and the other with a negative Seebeck coefficient in order to increase the generated voltage of the TC [111]. However, the sensitivity of a common TC is limited to 0.1 °C due to low electronic Seebeck coefficients [112]. Zhao et al. took advantage of the large ionic Seebeck coefficient found in polymer electrolytes and used it to gate a low-voltage organic transistor, which offers further signal amplification [113]. As shown in Fig. 12(a), one of the electrodes of the ionic thermoelectric device is connected to the gate of the electrolyte gated transistor [114], and the other electrode is grounded. Fig. 12(b) shows that temperature differences can modulate the source-drain current and Fig. 12(c) compares the transfer characteristic curves using temperature difference and applied voltage, and proves that the generated thermal voltage can be completely open and close the channel of the transistor. Interestingly, the combination of an ionic thermoelectric leg with a transistor forms a smart pixel, which is the basic circuit in an active matrix-addressed array. This is the first step towards temperature mapping. Note that the sensitivity can be improved simply by connecting a few couples of n-and p-legs to form an ionic thermoelectric module, the temperature sensitivity can be multiplied [68]. As shown in Fig. 12 (d and e), a 36-leg ionic thermopile connected to the gate of an electrolyte gated field-effect transistor can modulate the drain current by more than two orders of magnitude based on ΔT varying from −0.6 to 3.2 K [68]. The authors went further and proposed basic logic circuits, such as inverting a temperature signal input to an electrical signal output. This a potential avenue to use heat signals together with electrical signals, towards the vision of "thermo-electronics".

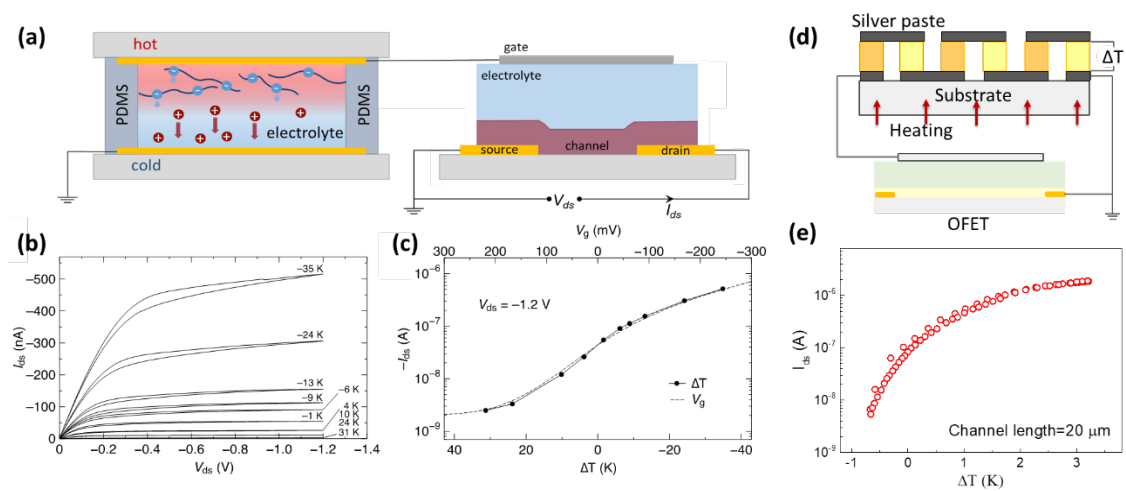

**Fig. 12**. Ionic thermoelectric gated transistor. (a) Illustration of the device structure. (b) Output characteristics at different fixed gating temperature differences (Ti gate, channel

length = 2 mm) (c) Transfer characteristics of the same transistor with temperature difference (solid dot) and applied gate potential (dashed line) [113]. (Reproduced with permission from Ref. [113], © Springer Nature 2017.) (d) Ionic thermopile composed of p and n types of legs connected in series. (e) Transfer characteristics of an electrolyte-gated transistor (channel length = 20 μm) gated by an ionic thermopile of 36 legs, when $\Delta T$ sweep from −0.67 K to 3.2 K [68]. (Reproduced with permission from Ref. [68], © Springer Nature 2019.)

Nowadays, temperature sensing electronics are crucial components in health monitoring systems for body condition assessment and disease diagnostics. Wearable and attachable temperature sensors can, e.g. provide information of core body temperature for people working in extreme environments, the fertility cycle of women, health condition of infants and wound healing conditions. Previous research has explored different heat sensing principles for this application, including heat-induced resistance changes of conducting materials [115], variations in the source-drain current of transistors [116], and frequency-changes in piezoelectric devices [117]. These sensors usually involve complicated and expensive microfabrication or patterning that are not compatible with large-scale production [118]. The utilization of ionic thermoelectric materials would lead to large output electric signal with simple and flexible structures. With the possibility to improve the manufacture methods, further miniaturization of the ionic thermopiles can boost the voltage responsivity by 100 times [113].

So far, the already developed good ionic thermoelectric materials range from liquid to solid, from water-based conductors to completely water-free, from good light absorber to transparent film. The large variation of the material properties enables multi-functionality of ionic thermoelectric materials. Fig. 13(a) shows an example, in which combining ionic Seebeck effect in mixed ion-electron conduction can enable the exclusive read-out of both temperature and humidity individually [118]. As shown in Fig. 13(b), the peak value of the voltage output (marked with a blue foreground) depends on both humidity and $\Delta T$. The ionic contribution to the Seebeck coefficient can be calculated and correlated to the humidity level through preliminary calibration.

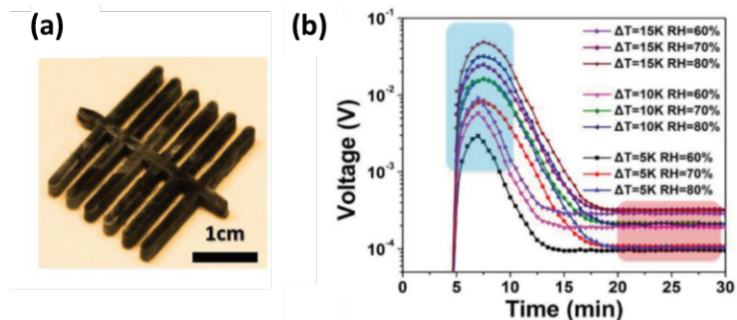

**Fig. 13**. (a) Photograph of a prepared mix-conductor aerogel. (b) Voltage axis intercept of *I-V* curves (thermal voltage output) as a function of time, under different humidity and temperatures [118]. (Reproduced with permission from Ref. [118], © Wiley-VCH 2019.)

Although the potential of applying ionic thermoelectric effect in temperature sensing has been demonstrated, there are few problems that need to be addressed before moving towards commercialization. The generated thermal voltage is proportional to temperature difference between two electrodes, if one electrode is exposed to constant temperature or heat dissipation, the absolute temperature of the other electrode can be obtained [113]. Otherwise, additional reference temperature is required in order to subtract the absolute temperature from the temperature difference. Still, even only tracking temperature difference between two interfaces over time (from $t1$ to $t2$) can directly provide information of the heat flux ($Q$) through the Eq. (15).

$$Q = \int_{t1}^{t2} \Delta T \kappa \mathrm{d}t \tag{15}$$

The cost-effective electrolytes will find important applications in industrial and domestic thermal insulation, as well as in high-tech instrument and products. Another concern is the generally relatively long responding time in ionic devices compared to that of electronic devices. Theoretically, the time needed to develop a concentration gradient induced by Soret effect is related to $L^2/D$ [119], in which $L$ is the distance between the cold and hot side and $D$ is the self-diffusion coefficient. From this relation, the responding time in an ionic thermoelectric device can be reduced by improving the ionic diffusion coefficient and reducing the thickness of the electrolytes. The ionic thermoelectric voltage can be combined with pyroelectric effect to enable fast response while keeping the stable response [120]. Theoretical studies also showing that the Soret coefficient could be accurately determined from the initial phase of the transient thermodiffusion [121,122].

Despite the advantages of the high output electric signal and material properties, the sensing mechanism in ionic thermoelectric materials is attractive for fundamental research. The procedure of ionic transport generating potential signal shares similarities with the thermosensation of mammals, which uses the temperature-sensitive transient receptor potential (thermoTRP) cation channels to convert the temperature change into electrical signals that are finally received as action potentials by nerve endings [123]. Early in 1956, the possibility of using ionic thermoelectric effect to mimic the thermal sensation of free-end nerve was proposed [124]. Recently, the transport of ions through nanochannels driven by temperature gradient has been reported to mimic the bio-thermosensation process [125,126]. The main limitation was the small ionic Seebeck coefficient of the aqueous solution (less than 1 mV K$^{-1}$). The emerging field of ionic thermoelectrics opens up new opportunities to explore and understand this process.

## 5. Conclusions and perspectives

As a recently renewed topic, ionic thermodiffusion has attracted the spotlight in thermoelectric studies. Since 2015, the field of ionic thermoelectrics is continuously striving to develop electrolytes with improved properties, which have enabled significant applications. However, the understanding of the underlying thermodynamics of the ionic thermodiffusion has not progressed accordingly. The difference of "heat of transport" between cation and anion in an electrolyte determines the sign and magnitude of the ionic Seebeck coefficient. At present, there is no satisfactory theoretical description for this parameter, not even in the seemingly simple case of a dilute electrolyte solution. Recent developments suggest that the giant values of Seebeck coefficient of polymer electrolytes and ions in solid matrices, could be related to their activated dynamics. Still, much remains to be done in order to predict and design ionic thermoelectric materials.

The finding that the thermodiffusion of ions can charge a supercapacitor or a battery brought this research area into thermoelectricity. The large ionic Seebeck coefficient offers electrolytes new opportunities not only in the areas of energy storage but also ultra-sensitive heat sensors and thermo-electronics. Meanwhile, the well-developed knowledge and technics in the fields of supercapacitors, micro-fabrication and characterization is ready to be incorporated with the new concepts. The electronic thermoelectric field is relatively mature with established theories and commercialized applications, while the journey of ionic thermoelectrics has just begun. An ionic thermoelectrically charged supercapacitor/batteries is an "all-in-one" device with the energy generation and storage in the same single device. It is electrically equivalent to an electronic thermoelectric generator charging a supercapacitor or a battery. The higher ionic Seebeck coefficient compared to electronic Seebeck coefficient enables the storage of much larger amount electricity in the "all-in-one" ionic thermoelectric device, since it scales with the square of the Seebeck coefficient. In practical applications, the feasibility for large-scale manufacturing, the cost of materials and processing and environmental impact are also critical measures. The mechanical properties of polymer ionic thermoelectrics, their mechanical softness or even stretchability makes them attractive for new applications, especially in wearable devices, functional band aids, sensors for packages.

**Acknowledgments**
D.Z. and X.C. would like to acknowledge Knut and Alice Wallenberg Foundation (Wallenberg Wood Science Center), the Swedish research council ("Ionic thermoelectric effect in electrolytes" 2018-04037 and "Next generation organic solar cells" 2016-06146), the Advanced Functional Materials Center at Linköping University (2009- 00971). A.W. acknowledges support by the French National Research Agency through grant ANR-19-CE30-0012 and by the European Research Council (ERC) through grant No. 772725.